\documentclass[journal, twocolumn]{IEEEtran}

\usepackage{amsmath, mathrsfs, graphicx,epsfig,amssymb, amsfonts}
\usepackage[center]{caption}

\newtheorem{lemma}{\bf Lemma}

\newtheorem{thm}{\bf Theorem}

\newtheorem{rem}{\bf Remark}
\newtheorem{cor}{\bf Corollary}
\newtheorem{ex}{\bf Example}

\begin{document}

\title{The diversity-multiplexing tradeoff of the symmetric MIMO $2$-user interference channel}

\author{{\Large Sanjay ~Karmakar ~~~~~~Mahesh~ K. ~Varanasi}
\thanks{S. Karmakar and M. K. Varanasi are both with the Department
of Electrical Computer and Energy Engineering, University of Colorado at Boulder, Boulder,
CO, 30809 USA e-mail: (sanjay.karmakar@colorado.edu, varanasi@colorado.edu).}
\thanks{This work was supported in part by the US National Science Foundation Grant CCF-0728955.}
}

%

\maketitle

\begin{abstract}
The fundamental diversity-multiplexing tradeoff (DMT) of the quasi-static fading, symmetric $2$-user MIMO interference channel (IC) with channel state information at the transmitters (CSIT) and a short term average power constraint is obtained. The general case is considered where the interference-to-noise ratio (INR) at each receiver scales differently from the signal-to-noise ratio (SNR) at the receivers. The achievability of the DMT is proved by showing that a simple Han-Kobayashi coding scheme can achieve a rate region which is within a constant (independent of SNR) number of bits from a set of upper bounds to the capacity region of the IC. In general, only part of the DMT curve with CSIT can be achieved by coding schemes which do not use any CSIT (No-CSIT). A result in this paper establishes a threshold for the INR beyond which the DMT with CSIT coincides with that with No-CSIT. Our result also settles one of the conjectures made in~\cite{EaOlCv}. Furthermore, the fundamental DMT of a class of non-symmetric ICs with No-CSIT is also obtained wherein the two receivers have different numbers of antennas.
\end{abstract}

%
\IEEEpeerreviewmaketitle

\section{Introduction}
The $2$-user IC is one of the most basic models of a general multiuser wireless network in which several transmit-receive
pairs communicate with each other in the face of interference. The $2$-user single-input single-output (SISO) IC being relatively better understood~\cite{ArPv}~\cite{WT}~\cite{ETW1} we consider in this paper the more general problem of characterizing the DMT of the $2$-user IC having multiple antennas at each node (MIMO). Depending on the number of antennas at different nodes, the SNRs and INRs at different nodes, and the availability of CSIT, the $2$-user MIMO IC can be divided into different classes. In this paper, MIMO ICs which have equal number of antennas at all nodes are called {\it symmetric} ICs and {\it asymmetric} ICs, otherwise. While some recent results \cite{ZG} and \cite{CvMv} point to a significant loss in degrees of freedom (DoF) or capacity pre-log factor -- which in turn implies a loss in DMT -- due to a complete lack of CSIT, the results of~\cite{EAKK} shows that the DMT with CSIT on a SISO IC can be achieved with only a single bit of feedback about the channel state. Anticipating that similar results can be found for MIMO ICs, the DMT for MIMO ICs with CSIT can be seen to be an important benchmark relative to which the performance of practical schemes with limited CSIT must be compared. On the other hand, the DMT without CSIT serves as a baseline from where marked improvements must be sought by an efficient use of limited CSIT. Finally, unlike the DMT framework in a point-to-point (PTP) channel~\cite{tse1}, where there is a single communication link which can be characterized by a single SNR, in a multiuser setting such as the one at hand, it is natural to allow the SNRs and INRs of different links to vary with a nominal SNR denoted as $\rho $ such that the ratio of the SNR or INR in dB to the nominal SNR in dB is held fixed as nominal SNR grows. This idea was introduced in~\cite{ETW1} leading to characterization of the so-called Generalized DoF (GDoF) region of the $2$-user SISO IC. Later, this technique was extended to the DMT scenario for the SISO IC in~\cite{ArPv},~\cite{ABarxiv}. Following a similar approach, we mathematically model the INRs at both the receivers as $\rho^\alpha$ for some $\alpha\geq 0$ and denote the SNRs at each receiver by $\rho$. We should refer to the corresponding DMT as the generalized DMT (GDMT) to distinguish it from the case when $\alpha=1$, i.e., SNR=INR in all the links but for simplicity we will henceforth refer to the GDMT simply as the DMT.

For the sake of simplicity in this paper, we characterize the DMT of a {\it symmetric} $2$-user MIMO IC with CSIT having $n$ antennas at each node. The DMT of the {\it asymmetric} MIMO IC with CSIT will be reported in~\cite{SkMv4}. We also characterize the DMT of a class of {\it asymmetric} MIMO IC with No-CSIT and $\alpha\geq 1$. To the best of our knowledge, this is the first result on the DMT of the {\it asymmetric} $2$-user MIMO IC with No-CSIT. In~\cite{EaOlCv} an upper bound to the usual DMT $(\alpha=1)$ of the {\it symmetric} $2$-user MIMO IC with CSIT was derived and conjectured to be tight. We prove this conjecture as a special case of the more general DMT result of this paper.

The rest of the paper is organized as follows. The channel model and the DMT notations in Section~\ref{sec_channel_model} are followed by the asymptotic joint eigenvalue distribution of three mutually correlated random matrices (correlated in a special way), which we derive in Section~\ref{subsec_eigenvalue_distribution}. In Section~\ref{subsec_approx_cap}, we derive a set of upper and lower bounds to the capacity region of the MIMO IC with CSIT. These bounds are then used to derive the DMT. In Section~\ref{sec_main_result}, we compute the explicit GDMT of the {\it symmetric} IC with CSIT. Finally, in Section~\ref{subsec_dmt_nocsit}, we characterize the DMT of a class of {\it asymmetric} MIMO ICs with No-CSIT.

{\bf Notation:} We will denote the conjugate transpose of the matrix $A$ by $A^{\dagger}$ and its determinant as $|A|$. $A\in \mathbb{C}^{n\times m}$ would mean that $A$ is a $n\times m$ matrix with entries in $\mathbb{C}$ where, $\mathbb{C}$  represents the field of complex numbers. The symbols $x\land y$, $x\vee y$ and $(x)^+$ represents the minimum and maximum between $x$ and $y$ and the maximum of $x$ and $0$, respectively. All the logarithms in this paper are with base $2$. We denote the distribution of a complex circularly symmetric Gaussian random vector with zero mean and covariance matrix $Q$, by $\mathcal{CN}(0,Q)$. If $\mathcal{R}$ represents a set of points in $\mathbb{R}^2$ then $\mathcal{R}\pm(c_1,c_2)=\{(R_1\pm c_1,R_2\pm c_2): (R_1,R_2)\in \mathcal{R} \}$.

\section{Channel Model and Preliminaries}
\label{sec_channel_model}
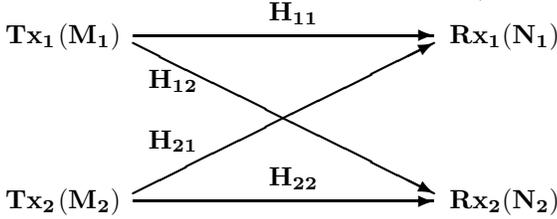
\begin{figure}[!thb]
\setlength{\unitlength}{1mm}
\begin{picture}(80,30)
\thicklines
\put(20,10){\vector(1,0){40}}
\put(20,32){\vector(1,0){40}}
\put(20,11){\vector(2,1){40}}
\put(20,31){\vector(2,-1){40}}
\put(10,9){$\mathbf{(M_2)}$}
\put(69,9){$\mathbf{(N_2)}$}
\put(10,31){$\mathbf{(M_1)}$}
\put(69,31){$\mathbf{(N_1)}$}

\put(3,9){$\mathbf{Tx_2}$}
\put(62,9){$\mathbf{Rx_2}$}
\put(3,31){$\mathbf{Tx_1}$}
\put(62,31){$\mathbf{Rx_1}$}

\put(38,12){$\mathbf{{H}_{22}}$}
\put(38,34){$\mathbf{{H}_{11}}$}
\put(22,17){$\mathbf{{H}_{21}}$}
\put(22,25){$\mathbf{{H}_{12}}$}
\end{picture}
\vspace{-5mm}
\caption{The 2-user MIMO interference channel.}
\label{channel_model_two_user_IC}
\end{figure}

Consider the MIMO IC shown in Figure~\ref{channel_model_two_user_IC}. Transmitter $1$ ($Tx_1$) and transmitter $2$ ($Tx_2$) have $M_1$ and $M_2$ antennas, respectively, and receiver $1$ ($Rx_1$) and receiver $2$ ($Rx_2$) have $N_1$ and $N_2$ antennas, respectively. Henceforth, such an IC will be referred to as an $(M_1,N_1,M_2,N_2)$ IC. ${H}_{ij}\in \mathbb{C}^{N_j\times M_i}$  is the  channel matrix between $Tx_i$ and $Rx_j$. $H_{ij}$ for $1\leq i,j\leq 2$ are mutually independent and contain mutually independent and identically distributed (i.i.d.) $\mathcal{CN}(0,1)$ entries. Following~\cite{ABarxiv}, we also incorporate a real-valued attenuation factor, denoted as $\eta_{ij}$, for the signal transmitted from $Tx_i$ to receiver $Rx_j$. At time $t$, $Tx_i$ chooses a vector ${X}_{it}\in \mathbb{C}^{M_i\times 1}$ and sends $\sqrt{P_i}{X}_{it}$ into the channel. The input signals are assume to satisfy the following short term average power constraint:

\begin{equation}
\label{power_constraint}
\textrm{tr}(Q_{it}) \leq M_i, \forall ~ i=1,2,~\textrm{where}~Q_{it}=\mathbb{E}\left({X}_{it}{X}_{it}^{\dagger}\right).
\end{equation}

With these aforementioned assumptions, the received signals at time $t$ can be written as
\begin{IEEEeqnarray*}{l}
\label{system_eq_two_user_IC1}
Y_{1t}=\eta_{11} \sqrt{P_1} {H}_{11}{X}_{1t}+ \eta_{21} \sqrt{P_2} {H}_{21}{X}_{2t}+Z_{1t},\\
Y_{2t}= \eta_{12} \sqrt{P_1} {H}_{12}{X}_{1t}+\eta_{22} \sqrt{P_2} {H}_{22}{X}_{2t}+Z_{2t},
\end{IEEEeqnarray*}
where $Z_{it}\in\mathbb{C}^{N_i\times 1}$ are i.i.d as $\mathcal{CN}(\mathbf{0}, I_{N_i})$ across $i$ and $t$. The above equations can be equivalently written in the following form:
\begin{IEEEeqnarray}{l}
\label{system_eq_two_user_IC2}
Y_{1t}=\sqrt{\textrm{SNR}_{11}}H_{11}\hat{X}_{1t}+ \sqrt{\textrm{INR}_{21}}H_{21}\hat{X}_{2t}+Z_{1t},\\
Y_{2t}= \sqrt{\textrm{INR}_{12}}H_{12}\hat{X}_{1t}+\sqrt{\textrm{SNR}_{22}}H_{22}\hat{X}_{2t}+Z_{2t},
\end{IEEEeqnarray}
where the normalized inputs $\hat{X}_i$'s satisfy equation \eqref{power_constraint} with equality and $\textrm{SNR}_{ii}$ and $\textrm{INR}_{ji}$ are the signal-to-noise ratio and interference-to-noise ratio at receiver $i$. In the analysis that follows, we will assume the following scaling parameters (with respect to a nominal SNR, $\rho$) for the different SNRs and INRs.
\begin{IEEEeqnarray}{l}
\alpha_{11}=\frac{\log (\textrm{SNR}_{11})}{\log(\rho)}=\alpha_{22}=\frac{\log(\textrm{SNR}_{22})}{\log(\rho)}=1,\\
~\alpha_{12}=\frac{\log(\textrm{INR}_{12})}{\log(\rho)}=\alpha_{21}=\frac{\log(\textrm{INR}_{21})}{\log(\rho)}=\alpha.
\end{IEEEeqnarray}
For ease of notation, we will henceforth set $\textrm{SNR}_{ii}=\rho_{ii}$, $\textrm{INR}_{ij}=\rho_{ij}$, $H=\{H_{ij}, 1\leq i,j\leq 2\}$ and $\bar{\rho}=\{\rho_{ij},1\leq i,j\leq 2\}$.

To define the DMT notation we follow~\cite{tse1}. We assume that user $i$ uses a coding scheme $\mathscr{C}_i$ and is operating at a rate $R_i=r_1 \log(\rho)$ bits per channel use. Let us denote $\mathscr{C}=\{\mathscr{C}_1,\mathscr{C}_2\}$. The diversity order of the IC with coding scheme $\mathscr{C}$ and rates $(R_1,R_2)$ is defined as
\begin{IEEEeqnarray}{l}
\label{def_achievable_dmt}
d_{IC}(r_1,r_2,\mathscr{C})=\lim_{\rho \to \infty} -\frac{\log \left(P_e(\bar{\rho})\right)}{\log(\rho)},
\end{IEEEeqnarray}
where $P_e(\bar{\rho})$ $=(P_{e_1}(\bar{\rho})\vee P_{e_2}(\bar{\rho}))$ with $P_{e_i}(\bar{\rho})$ denoting the probability of error (averaged over channel statistics) at receiver $i$. Finally, the fundamental DMT (henceforth, just DMT) of the IC, denoted as $d_{IC}^*(r_1,r_2)$, is defined as
\begin{eqnarray}
\label{def_opt_dmt}
d_{IC}^*(r_1,r_2)= \max_{\mathscr{C}\in \widetilde{\mathscr{C}}} ~d_{IC}(r_1,r_2,\mathscr{C}),
\end{eqnarray}
where $\widetilde{\mathscr{C}}$ represents the collection of all coding schemes that use CSIT and the short term power constraint (equation \eqref{power_constraint}). Note the diversity order $d_{IC}^*(r_1,r_2)$ is a function of the relative scaling parameters of the different links ($\alpha$). However, for brevity, we will not mention them explicitly.

\subsection{Asymptotic Eigenvalue Distribution}
\label{subsec_eigenvalue_distribution}
In this subsection, we will derive the joint distribution of the scaling parameters of the eigenvalues of three correlated random Wishart matrices, which will be used later in section~\ref{sec_main_result} to derive the DMT of the MIMO IC.

\begin{figure*}[!t]
\begin{IEEEeqnarray}{l}
\label{eq_pdf_exponent}
\mathcal{E}(\mathcal{S})= \sum_{j=1}^{n}\left((2n+1-2j)\beta_j-n(\alpha-\alpha_j)^+-n(\alpha-\gamma_j)^+ +\sum_{i=1}^{(n-j)}\max \{\alpha-\beta_j-((\alpha_i\land \gamma_i)\land (\gamma_i+\alpha_i-\alpha))
\}^+\right)
\end{IEEEeqnarray}
\hrulefill
\end{figure*}

\begin{thm}
\label{thm_eigen_value_distribution}
Let $H_i\in \mathbb{C}^{n\times n}$ for $i\in \{1,2,3\}$ are three mutually independent random matrices with i.i.d. $\mathcal{CN}(0,1)$ entries and $x_1\geq \cdots \geq x_n>0$, $\mu_1\geq \cdots \geq \mu_n >0$ and $\lambda_1\geq \cdots \geq \lambda_n >0$ be the ordered non-zero eigen-values of $W_1=\tilde{H}_1\tilde{H}_1^\dagger$, $W_2=H_2 H_2^{\dagger}$ and $W_3=H_3^{\dagger}H_3$, respectively, where $\tilde{H}_1=\left(I_n+\rho^{\alpha} H_2H_2^{\dagger}\right)^{-\frac{1}{2}}H_1\left(I_n+\rho^\alpha H_3^\dagger H_3\right)^{-\frac{1}{2}}$. Further, if we assume that $x_i=\rho^{-\beta_i}, \mu_i=\rho^{-\gamma_i}$ and $\lambda_i=\rho^{-\alpha_i}, ~\forall 1\leq i\leq n $ and $\rho \to \infty (\rho\in \mathbb{R}^+)$, then the joint distribution of $\vec{\beta}$ given $\vec{\gamma}$ and $\vec{\alpha}$ (with $\vec{\beta}=\{\beta_1, \cdots ,\beta_n\}$ with $\vec{\gamma}$ and $\vec{\alpha}$ similarly defined) is given as
\begin{equation*}
f(\vec{\beta}|\vec{\gamma},\vec{\alpha})\dot{=}
\left\{\begin{array}{l}
\rho^{-\mathcal{E}(\mathcal{S})},~\textrm{if}~ (\vec{\beta},\vec{\gamma},\vec{\alpha}) \in \mathscr{B};\\
0, ~\textrm{otherwise},
\end{array}\right.
\end{equation*}
where $\mathcal{S}=\{\vec{\beta},\vec{\gamma},\vec{\alpha},\alpha\}$, $\mathcal{E}(\mathcal{S})$ is given by equation \eqref{eq_pdf_exponent} and
\begin{IEEEeqnarray*}{l}
\mathscr{B}=\Big\{(\vec{\beta},\vec{\gamma},\vec{\alpha}):\beta_1\geq 0, \alpha_i+\beta_j\geq \alpha,  ~\textrm{and }\\
\quad \quad \quad \quad \quad \quad \gamma_i+\beta_j\geq \alpha, ~\forall (i+j)\geq (n+1)\Big\}.
\end{IEEEeqnarray*}
\end{thm}
\begin{rem}
Since $W_2$ and $W_3$ are independent so are $\vec{\gamma}$ and $\vec{\alpha}$. Now importing the distributions of $\vec{\gamma}$ and $\vec{\alpha}$ from~\cite{tse1} and using it in
\begin{equation*}
f(\vec{\beta},\vec{\gamma},\vec{\alpha})=f(\vec{\beta}|\vec{\gamma},\vec{\alpha})f(\vec{\gamma},\vec{\alpha})=f(\vec{\beta}|\vec{\gamma},\vec{\alpha})f(\vec{\gamma})f(\vec{\alpha}),
\end{equation*}
the joint distribution of $(\vec{\beta},\vec{\gamma},\vec{\alpha})$ can be derived. Further the above theorem can be generalized to the case of non-square $H_i$s, which is necessary to derive the DMT of an IC with arbitrary number of antennas at each node (this will be done in \cite{SkMv4}).
\end{rem}

\subsection{Approximate capacity region}
\label{subsec_approx_cap}
In this subsection, we will find an upper and a lower bound to the capacity region of the $2$-user MIMO IC, which in the next subsection will be used to derive DMT upper and lower bounds, respectively. We start with the upper bound.
\begin{lemma}
\label{lem_upper_bound}
For the $2$-user MIMO IC shown in Figure~\ref{channel_model_two_user_IC} and given realization of channel matrices $H$, the capacity region is contained in the following set of rate tuples
\begin{IEEEeqnarray*}{l}
\mathcal{R}^c(H,\bar{\rho})+(N_1\log(M_1\vee M_2),N_2\log(M_1\vee M_2)),
\end{IEEEeqnarray*}
where $\mathcal{R}^c(H,\bar{\rho})$ represents the set of rate pairs $(R_1,R_2)$ such that $R_1,R_2\geq 0$ and
\begin{IEEEeqnarray*}{l}
\label{eq_bound1}
R_i\leq \log \left| \left(I_{N_i}+\rho_{ii} H_{ii}H_{ii}^{\dagger}\right)\right|\triangleq I_{bi},~\textrm{for}~i\in{1,2};\\
R_1+R_2\leq \log \left| \left(I_{N_2}+\rho_{12} H_{12}H_{12}^{\dagger}+\rho_{22} H_{22}H_{22}^{\dagger}\right)\right|+ \nonumber \\
\label{eq_bound3}
\quad \log \left| \left(I_{M_1}+\rho_{11} H_{11}P_{12}^{-1}H_{11}^{\dagger}\right)\right|\triangleq I_{b3};\\
R_1+R_2\leq \log \left| \left(I_{N_1}+\rho_{21} H_{21}H_{21}^{\dagger}+\rho_{11}  H_{11}H_{11}^{\dagger}\right)\right| \\
\label{eq_bound4}
 \quad \log \left| \left(I_{M_2}+\rho_{22}  H_{22} P_{21}^{-1}H_{22}^{\dagger}\right)\right| \triangleq I_{b4};\\
R_1+R_2 \leq \log \left| \Big( I_{N_1}+\rho_{11} H_{11} P_{12}^{-1}H_{11}^{\dagger}+\rho_{21} H_{21}H_{21}^{\dagger}\Big)\right| \nonumber \\
\label{eq_bound5}
\quad + \log \left| \Big(I_{N_2}+\rho_{12} H_{12}H_{12}^{\dagger}+ \rho_{22} H_{22} P_{21}^{-1}H_{22}^{\dagger}\Big)\right|\triangleq I_{b5};\\
2R_1+R_2\leq \log \left| \left(I_{N_1}+\rho_{21}  H_{21}H_{21}^{\dagger}+ \rho_{11} H_{11}H_{11}^{\dagger}\right)\right|+ \nonumber \\
\quad \log \left| \left(I_{N_2}+ \rho_{12}  H_{12}H_{12}^{\dagger}+ \rho_{22} H_{22} P_{21}^{-1}H_{22}^{\dagger}\right)\right|\nonumber\\
\label{eq_bound6}
\quad \quad +\log \left| \left(I_{N_1}+\rho_{11} H_{11} P_{12}^{-1}H_{11}^{\dagger}\right)\right|\triangleq I_{b6}; \\
R_1+2R_2\leq \log \left| \left(I_{N_2}+\rho_{12}  H_{12}H_{12}^{\dagger}+ \rho_{22} H_{22}H_{22}^{\dagger}\right)\right|+ \nonumber \\
\quad \log \left| \left(I_{N_1}+ \rho_{21} H_{21}H_{21}^{\dagger}+\rho_{11} H_{11} P_{12}^{-1}H_{11}^{\dagger}\right)\right|\nonumber\\
\label{eq_bound7}
\quad \quad + \log\left|\left(I_{N_2}+\rho_{22} H_{22} P_{21}^{-1}H_{22}^{\dagger}\right)\right|\triangleq I_{b7},
\end{IEEEeqnarray*}
where $P_{ij}=\left(I_{M_i}+\rho_{ij} H_{ij}^{\dagger}H_{ij}\right)$ for $i\neq j\in \{1,2\}$.
\end{lemma}

\begin{rem}
\label{rem_incorrect_bound_leveque}
Note a similar set of upper bounds on the capacity region was also derived in~\cite{EaOlCv}. We see that the first four bounds in Lemma~\ref{lem_upper_bound} are identical to those in~\cite{EaOlCv} and the fifth bound can be shown to be equivalent. However, the last two bounds are different. It should also be noticed that while specialized to case $M_1=N_1=M_2=N_2=1$, the last two bounds of~\cite{EaOlCv} do not match with the corresponding bounds of\cite{ETW1} whereas, all the bounds of Lemma~\ref{lem_upper_bound} do.
\end{rem}

Next we find the achievable rate region of a simple Han-Kobayashi~\cite{HK} coding scheme. Suppose each user's message is divided into two parts (private and public, respectively) and is encoded using a random Gaussian code. Thus the codewords can be written as
\begin{equation}
X_1=U_1+W_1 ~\textrm{and}~
X_2=U_2+W_2,
\end{equation}
where $U_1, W_1$ and $U_2, W_2$ are (the private and public parts of the messages of $Tx_1$ and $Tx_2$, respectively) mutually independent complex Gaussian random vectors using the following channel dependent power split (note this power split satisfies the power constraint in \eqref{power_constraint}) among the private and common parts: $K_{i1}=\mathbb{E}(W_iW_i^{\dagger})$ and $K_{i2}=\mathbb{E}(U_iU_i^{\dagger})$, where
\begin{IEEEeqnarray}{c}
\label{eq_power_split}
K_{i1}=\frac{I_{M_i}}{2}~\textrm{and}~
K_{i2}=\frac{1 }{2}\left(I_{M_i}+\rho_{ij} H_{ij}^{\dagger}H_{ij}\right)^{-1}.
\end{IEEEeqnarray}
We refer to this coding scheme as $\mathcal{HK}(K_{11},K_{12},K_{21},K_{22})$ scheme.

\begin{lemma}
\label{lem_achievable_region}
For a given channel realization $H$, the $\mathcal{HK}(K_{11},K_{12},K_{21},K_{22})$ scheme, where $K_{ij},~1\leq i,j\leq 2$ is given by \eqref{eq_power_split}, can achieve all the rate pairs $(R_1,R_2)$ such that $(R_1,R_2)\in \{\mathcal{R}^c\left(H,\bar{\rho}\right)-(2N_1,2N_2)\}$, where $\mathcal{R}^c\left(H,\bar{\rho}\right)$ is as given in Lemma~\ref{lem_upper_bound}.
\end{lemma}

Using Lemma~\ref{lem_upper_bound} and \ref{lem_achievable_region}, respectively and similar method as in the proof of Theorem~$2$ in~\cite{tse1} it can be proved that (for more details refer to~\cite{SkMv4})
\begin{IEEEeqnarray}{rl}
\label{eq_dmt_exp1}
d_{IC}^*(r_1,r_2)=& \min_{i\in \mathcal{I}}{d_{O_i}({r}_i)},\\
\label{eq_outage_probability}
\textrm{where}~\rho^{-d_{O_i}({r}_i)}\dot{=}& \Pr\left(I_{bi}\leq r_i\right),
\end{IEEEeqnarray}
for $i\in \mathcal{I}=\{1,\cdots ,7\}$ and $r_3=r_4=r_5=(r_1+r_2)$, $r_6=(2r_1+r_2)$ and $r_7=(r_1+2r_2)$.

\section{Explicit DMT of the $(n,n,n,n)$ IC}
\label{sec_main_result}
\label{sec_DMT_Z_channel}
In this section we will evaluate $d_{O_i}(r_i)$'s given in equation \eqref{eq_dmt_exp1} which would yield the explicit DMT expressions for the IC. Using the first and second bound of Lemma~\ref{lem_upper_bound} in equations~\eqref{eq_outage_probability} it can be proved that
\begin{IEEEeqnarray}{l}
\label{eq_do1}
d_{O_i}(r_i)=d_{n,n}\left(r_i\right),~\forall ~ r_i\in [0,~ n],
\end{IEEEeqnarray}
where $d_{p,q}(r)$ is the optimal diversity order of a point-to-point (PTP) MIMO channel with $p$ transmit and $q$ receive antennas and $i\in \{1,2\}$. To evaluate $d_{O_3}(r_3)$, we write  the bound $I_{b3}$ of Lemma~\ref{lem_upper_bound} in the following way
\begin{IEEEeqnarray*}{rl}
I_{b3}=&\log \left| \left(I_n+\rho \widetilde{H}_{11}^{\dagger}\widetilde{H}_{11}\right)\right|
+\log \left| \left(I_n+\rho  \widetilde{H}_{22} \widetilde{H}_{22}^{\dagger}\right)\right|\\
& + \log \left|\left(I_n+\rho^{\alpha} H_{21}H_{21}^{\dagger}\right)\right|, ~\textrm{where}
\end{IEEEeqnarray*}
$\scriptstyle{\widetilde{H}_{11}=\Big(\rho^\alpha H_{21}H_{21}^{\dagger}+I_n\Big)^{-\frac{1}{2}}H_{11}}$ and $\scriptstyle{\widetilde{H}_{22}=H_{22} \left(I_n+\rho^\alpha  H_{21}^{\dagger}H_{21}\right)^{-\frac{1}{2}}}$. To compute $d_{O_3}$ we need the asymptotic joint distributions of the eigen-values of mutually correlated matrices $\widetilde{H}_{11}^\dagger \widetilde{H}_{11}, \widetilde{H}_{22} \widetilde{H}_{22}^\dagger$ and $H_{21} H_{21}^\dagger$. This joint distribution can be derived using Theorem~$1$ of \cite{SkMv_asilomar_DDF}. Now, following a similar approach as in~\cite{tse1}, $d_{O_3}(r_s)$ can be evaluated.
\begin{IEEEeqnarray}{l}
\label{eq_do34a}
\textrm{For}~ \alpha \leq 1,~d_{O_3}(r_s)=\nonumber\\
\left\{\begin{array}{l}
\alpha d_{n,3n}(\frac{r_s}{\alpha})+2n^2(1-\alpha), ~\textrm{for}~ 0\leq r_s \leq n\alpha;\\
2(1-\alpha) d_{n,n}(\frac{(r_s-n\alpha)}{2(1-\alpha)}), ~\textrm{for}~ n\alpha\leq r_s \leq n(2-\alpha);
\end{array}\right.\end{IEEEeqnarray}
and for  $\alpha\geq 1 ,~ d_{O_3}(r_s)=$
\begin{IEEEeqnarray}{l}
\label{eq_do34b}
\left\{\begin{array}{l}
d_{n,3n}(r_s)+n^2(\alpha-1), ~\textrm{for}~ 0\leq r_s \leq n;\\
(\alpha-1) d_{n,n}(\frac{(r_s-n)}{(\alpha-1)}), ~\textrm{for}~ n\leq r_s \leq n\alpha.
\end{array}\right.
\end{IEEEeqnarray}
where $r_s=(r_1+r_2)$. Also, from symmetry we have $d_{O_4}(r_s)=d_{O_3}(r_s)$. To evaluate $d_{O_5}(r_s)$, we write  the bound $I_{b5}$ of Lemma~\ref{lem_upper_bound} in the following way
\begin{IEEEeqnarray*}{l}
I_{b5}=\log \left| \left(I_n+\rho \widetilde{H}_{11}\widetilde{H}_{11}^{\dagger}\right)\right| +\log \left| \left(I_n+\rho  \widetilde{H}_{22} \widetilde{H}_{22}^{\dagger}\right)\right|\\
\quad + \log \left|\left(I_n+\rho^\alpha H_{21}H_{21}^{\dagger}\right)\right|+\log \left|\left(I_n+\rho^\alpha H_{12}H_{12}^{\dagger}\right)\right|,
\end{IEEEeqnarray*}
with $\scriptstyle{\widetilde{H}_{ii}=\left(I_n+\rho^\alpha H_{ji}H_{ji}^{\dagger}\right)^{-\frac{1}{2}}H_{ii}\left(I_n+\rho^\alpha H_{ij}^{\dagger}H_{ij}\right)^{-\frac{1}{2}}}$. Noting that, given the eigenvalues of $H_{12}H_{12}^{\dagger}$ and $H_{21}H_{21}^{\dagger}$ the eigenvalues of $\widetilde{H}_{11}\widetilde{H}_{11}^{\dagger}$ and $\widetilde{H}_{22}\widetilde{H}_{22}^{\dagger}$ are independent, we can use Theorem~\ref{thm_eigen_value_distribution} to find the joint distribution of the eigenvalues of $\widetilde{H}_{11}\widetilde{H}_{11}^{\dagger}, \widetilde{H}_{22}\widetilde{H}_{22}^{\dagger}, H_{12}H_{12}^{\dagger}$ and $H_{21}H_{21}^{\dagger}$ (refer to~\cite{SkMv4} for more details). Using this distribution result, equation \eqref{eq_outage_probability} can be evaluated for $d_{O_5}(r_s)$ as follows.
\begin{IEEEeqnarray}{l}
\label{eq_do5a}
\textrm{For}~ \alpha \leq \frac{1}{2},~d_{O_5}(r_s)=\nonumber\\
\left\{\begin{array}{l}
2 \alpha d_{n,3n}\left(\frac{r_s}{2\alpha}\right)+2n^2(1-2\alpha), ~\textrm{for}~ 0\leq r_s \leq 2n\alpha;\\
2(1-2\alpha) d_{n,n}\left(\frac{(r_s-2n\alpha)}{2(1-2\alpha)}\right), ~\textrm{for}~ 2n\alpha\leq r_s \leq 2n(1-\alpha);
\end{array}\right.\\
\label{eq_do5b}
\textrm{and for}~ \frac{1}{2}\leq \alpha ,~d_{O_5}(r_s)=\nonumber\\
\left\{\begin{array}{l}
n^2(2\alpha-1)+ d_{n,3n}(r_s), ~\textrm{for}~0\leq r_s\leq n;\\
(2\alpha-1)d_{n,n}\left(\frac{(r_s-n)}{(2\alpha-1)}\right),~\textrm{for}~n\leq r_s\leq 2n\alpha.
\end{array}\right.
\end{IEEEeqnarray}
Using a similar approach, $d_{O_6}(r_t)$ can also be derived. For even $n$ it is given as
\begin{IEEEeqnarray}{l}
\textrm{For}~ \alpha \leq \frac{1}{2},~d_{O_6}(r_t)=\nonumber\\
\label{eq_do6a}
\left\{\begin{array}{l}
n^2(2-\alpha)+\alpha d_{n,3n}\left(\frac{r_t}{\alpha}\right), ~ 0\leq r_t \leq \frac{n\alpha}{2};\\
n^2(2-3\alpha)+\alpha d_{n,3n}\left(\frac{r_t^1}{2\alpha}+\frac{n}{2}\right)+\alpha d_{n,2n}\left(\frac{r_t^1}{2\alpha}\right), \\
~~~~\textrm{for}~0\leq r_t^1=r_t-\frac{n\alpha}{2} \leq n\alpha;\\
\scriptstyle{n^2(1-\alpha)+(1-2\alpha) d_{n,n}\left(\frac{r_t^2}{(1-\alpha)}\right)+\alpha d_{n,2n}\left(\frac{r_t^2}{(1-\alpha)}+\frac{n}{2}\right)},\\
~~~~\textrm{for}~ 0\leq r_t^2=r_t-\frac{3n\alpha}{2} \leq \frac{n(1-\alpha)}{2};\\
(1-2\alpha) d_{n,n}\left(\frac{r_t^3}{(3-4\alpha)}+\frac{n}{2}\right)+(1-\alpha)d_{n,n}\left(\frac{r_t^3}{(3-4\alpha)}\right),\\
~~~~\textrm{for}~0\leq r_t^3=r_t-\frac{n(1+2\alpha)}{2} \leq \frac{n(3-4\alpha)}{2};\\
(1-\alpha) d_{n,n}\left(\frac{(r_t-n)}{2(1-\alpha)}\right),~n(2-\alpha)\leq r_t\leq n(3-2\alpha),
\end{array}\right.\\
\textrm{and for}~ \frac{1}{2}\leq \alpha \leq 1,~d_{O_6}(r_t)=\nonumber\\
\label{eq_do6b}
\left\{\begin{array}{l}
n^2(2-\alpha)+\alpha d_{n,3n}\left(\frac{r_t}{\alpha}\right), ~ 0\leq r_t \leq \frac{n\alpha}{2};\\
n^2\alpha +\alpha d_{n,3n}\left(r_t^1+\frac{n}{2}\right)+(1-\alpha) d_{n,2n}\left(r_t^1\right), \\
~~~~\textrm{for}~0\leq r_t^1=r_t-\frac{n}{2} \leq n\alpha;\\
\scriptstyle{n^2(1-\alpha)+(2\alpha-1) d_{n,n}\left(\frac{r_t^2}{\alpha}\right)+(1-\alpha) d_{n,2n}\left(\frac{r_t^2}{\alpha}+\frac{n}{2}\right)},\\
~~~~\textrm{for}~ 0\leq r_t^2=r_t-\frac{n(\alpha+1)}{2} \leq \frac{n\alpha}{2};\\
(2\alpha-1) d_{n,n}\left(r_t^3+\frac{n}{2}\right)+(1-\alpha)d_{n,n}\left(r_t^3\right),\\
~~~~\textrm{for}~0\leq r_t^3=r_t-\frac{n(1+2\alpha)}{2} \leq \frac{n}{2};\\
(1-\alpha) d_{n,n}\left(\frac{(r_t-2n\alpha)}{2(1-\alpha)}\right),~n(1+\alpha)\leq r_t\leq 2n.
\end{array}\right.
\end{IEEEeqnarray}
\begin{IEEEeqnarray}{l}
\textrm{and for}~ 1\leq \alpha ,~d_{O_6}(r_t)=\nonumber \\
\label{eq_do6c}
\left\{\begin{array}{l}
n^2(2\alpha-1)+ d_{n,3n}(r_t), ~\textrm{for}~0\leq r_t\leq n;\\
(2\alpha-1)d_{n,n}\left(\frac{(r_t-n)}{(2\alpha-1)}\right),~\textrm{for}~n\leq r_t\leq 2n\alpha.
\end{array}\right.
\end{IEEEeqnarray}
where $r_t=(2r_1+r_2)$. From symmetry, we have $d_{O_6}(r_1+2r_2)=d_{O_7}(r_1+2r_2)$. Analytical expressions for $d_{O_6}$ for odd $n$ can be similarly derived and are not given here due to space constraints.

\begin{rem}
It should be noted from equations \eqref{eq_do5b} and \eqref{eq_do6c} that, on a $(n,n,n,n)$ MIMO IC, for $\alpha\geq 1$, $d_{O_6}$ ($d_{O_7}$) provides a strictly tighter bound on the optimal DMT than $d_{O_5}$ if $r_1\neq 0$ ($r_2\neq 0$).
\end{rem}

\begin{thm}
The optimal diversity order, $d_{IC}^*(r_1,r_2)$ at a multiplexing gain pair $(r_1,r_2)$, of a $(n,n,n,n)$ MIMO IC (Figure~\ref{channel_model_two_user_IC}), with CSIT, short term average power constraint, \eqref{power_constraint} and $\alpha_{11}=\alpha_{22}=1$, $\alpha_{12}=\alpha_{21}=\alpha \geq 0$, is given as
\begin{equation*}
d_{IC}^*(r_1,r_2)= \min_{1\leq i\leq 7} \{d_{O_i}(r_i)\},
\end{equation*}
where $r_3=r_4=r_5=(r_1+r_2)$, $r_6=(2r_1+r_2)$ and $r_7=(r_1+2r_2)$ and $d_{O_i}$s are given by equations \eqref{eq_do1}-\eqref{eq_do6c}.
\end{thm}

\begin{cor}
\label{cor1}
The optimal DMT of a $(n,n,n,n)$ IC, with CSIT and $\alpha_{ij}=1,~\forall ~i,j\in\{1,2\}$, at a MG tuple $(r_1,r_2)$ is given as
\begin{equation*}
d_{IC}^*(r_1,r_2)=
\min \{d_{n,n}(r_1),d_{n,n}(r_2),d_{n,3n}(r_1+r_2)\}.
\end{equation*}
\end{cor}

\begin{figure}[!hbt]
\centering
\includegraphics[width=8.0cm,height=5cm,]{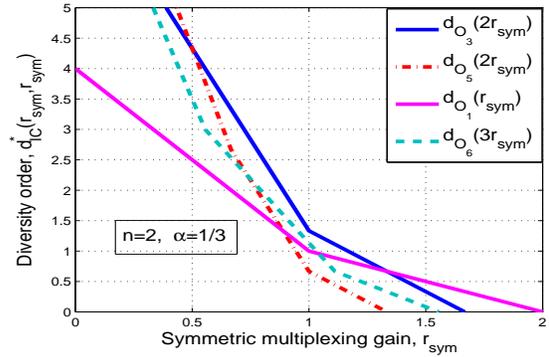}
\caption{Different DMT bounds on a MIMO IC.}
\label{figure_symmetric_IC_n2_a1by3}
\end{figure}
\begin{rem}
Corollary \ref{cor1} follows from the fact that for $\alpha=1$, the $3^{rd}$ bound of Lemma~\ref{lem_upper_bound} is tighter than the $5^{th}, 6^{th}$ or $7^{th}$ in the special case considered. In fact, this was the reason based on which the authors in~\cite{EaOlCv} conjectured the result of Corollary~\ref{cor1}. However, the fact that this is not true in general, i.e., for $\alpha\neq 1$, is illustrated in Figure~\ref{figure_symmetric_IC_n2_a1by3}, where we have plotted the outage exponents corresponding to all the bounds of Lemma~\ref{lem_upper_bound} on a $(2,2,2,2)$ IC with $\alpha=\frac{1}{3}$. Contrary to the case when $\alpha=1$, in this case at high MGs the $5^{th}$ bound is tighter than the $3^{rd}$.
\end{rem}

\begin{figure}[!hbt]
\centering
\includegraphics[width=8.0cm,height=5cm,]{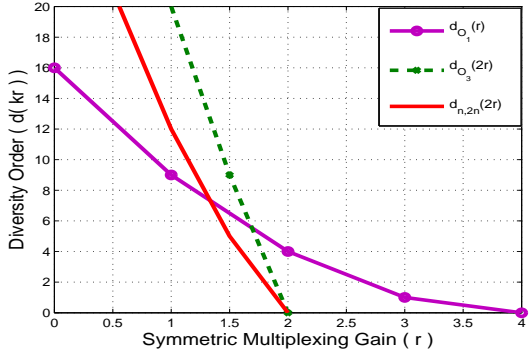}
\caption{DMT on a $(4,4,4,4)$ IC with and without CSIT.}
\label{figure_IC_example1}
\end{figure}

\begin{ex}
Consider an $(4,4,4,4)$ IC, with $\alpha_{ij}=1$ for all $1\leq i,j\leq 2$ and $r_1=r_2=r$. In Figure~\ref{figure_IC_example1}, $\min \{d_{O_1}(r), d_{O_3}(2r)\}$ represents the optimal DMT with CSIT (Corollary~\ref{cor1}) and $\min \{d_{O_1}(r), d_{n,2n}(2r)\}$ represents an achievable DMT when both the sources treat the channel to each receiver as a multiple-access channel (MAC) and use channel independent Gaussian codes. We see that at low MGs, the fundamental DMT with CSIT can be achieved with no CSIT at all. In the following subsection we shall show that, for some antenna configurations this is true for all MGs.
\end{ex}

\vspace{-5mm}
It can be proved that, on an $(n,n,n,n)$ IC with $\alpha\geq 1$ and $r_1=r_2=r$, the $\mathcal{HK}\left(I_n,0_n,I_n,0_n\right)$ scheme can achieve the following DMT (Lemma~$4$,~\cite{SKMV_Z_channel_DMT_conference})
\begin{IEEEeqnarray*}{l}
d_{MAC}(r)=\min \left\{d_{n,n}(r),~d_s(2r)\right\},~\textrm{where}\\
d_s(2r)=\left\{\begin{array}{c}
d_{n,3n}(2r)+n^2(\alpha-1), ~\textrm{for}~ 0\leq 2r \leq n;\\
(\alpha-1) d_{n,n}(\frac{(2r-n)}{(\alpha-1)}), ~\textrm{for}~ n\leq 2r \leq n\alpha.
\end{array}\right.
\end{IEEEeqnarray*}
Comparing this with equation \eqref{eq_do34b} we have
\begin{thm}
The optimal diversity order, $d_{\textrm{IC}_1}^*(r)$ at a multiplexing gain pair $(r,r)$, of a $(n,n,n,n)$ MIMO IC, with CSIT, short term average power constraint, \eqref{power_constraint} and $\alpha_{11}=\alpha_{22}=1$, $\alpha_{12}=\alpha_{21}=\alpha \geq 1.25$, can be achieved by the $\mathcal{HK}\left(I_n,0_n,I_n,0_n\right)$ scheme, i.e., $
d_{\textrm{IC}_1}^*(r)= d_{MAC}(r)$.
\end{thm}
Note that $\mathcal{HK}\left(I_n,0_n,I_n,0_n\right)$ scheme does not require any CSIT. In the next section we consider asymmetric ICs where the fundamental DMT with CSIT coincides with that with No-CSIT.

\subsection{DMT of an Asymmetric IC with No-CSIT}
\label{subsec_dmt_nocsit}
In this subsection, we derive the optimal DMT of the $2$-user MIMO IC with No-CSIT for a particular antenna configuration. In equations \eqref{eq_do34a}-\eqref{eq_do34b}, $d_{O_3}(r_3)$ was evaluated for $M_i=N_i=n$. However, note that the distribution results in~\cite{SkMv_asilomar_DDF} are valid for arbitrary $M_i$ and $N_i$ and can be used to evaluate $d_{O_3}(r_3)$. Also $d_{O_3}(r_3)$, being an upper bound (equation \eqref{eq_dmt_exp1}) to the optimal DMT of an IC with CSIT, is also an upper bound to the optimal DMT of the corresponding IC with No-CSIT. However, it can be proved that a joint maximum likelihood (ML) decoder at both the receivers can achieve $d_{O_3}(r_3)$ if both the users use random Gaussian codes (with identity as the covariance matrices), and when $M_1=M_2=M$, $2M\leq N_1 \leq N_2$ and $1\leq \alpha$.

\begin{lemma}
\label{lem_DMT_MleqN_case}
Consider the MIMO IC, as shown in Figure~\ref{channel_model_two_user_IC} with, $M_1=M_2=M$, $2M \leq N_1\leq N_2$ and $1\leq \alpha$. The optimal DMT of this channel with No-CSIT, at multiplexing gain pair $(r_1,r_2)$, is given by
\begin{IEEEeqnarray*}{l}
d_{IC,No-CSIT}^*(r_1,r_2)=\min \left\{d_{M,N_i}(r_i), d_{IC_s}(r_1+r_2)\right\}
\end{IEEEeqnarray*}
where $i\in\{1,2\}$ and for $k\in \{0,1\cdots (M-1)\}$,
\begin{IEEEeqnarray*}{l}
d_{IC_s}(r_s)=\\
\left\{\begin{array}{l}
\alpha d_{M,(M+N_1)}(\frac{r_s}{\alpha})+M((r_s-k\alpha-1)^+\\
+(M-k)(1-\alpha))+M(N_1-M), ~\forall r_s\in [k\alpha, (k+1)\alpha];\\
d_{M,(N_1-M)}(r_s-M\alpha), ~\forall r_s\in [M\alpha, M(\alpha-1)+N_1].
\end{array}\right.
\end{IEEEeqnarray*}
\end{lemma}
Note the optimal DMT for the case when $2M\leq N_2\leq N_1$ can be similarly found.

\section{Conclusion}
The fundamental DMT of the MIMO IC, with CSIT is characterized. In general it is an upper bound for the No-CSIT DMT of a corresponding MIMO IC. One class of ICs is identified for which the DMTs with and without CSIT coincide. However, finding all such MIMO ICs for which this happens is an interesting open problem. It is shown that in the DMT optimal scheme with full CSIT, a transmitter does not utilize the channel information of the direct link at all but fully uses the channel information of the cross link. Finding the minimum amount of channel information which is sufficient to achieve the full CSIT DMT is another interesting open problem.

\bibliographystyle{IEEEtran}
\bibliography{mybibliography}

\end{document}